\documentclass[aps,prl,twocolumn,showpacs,superscriptaddress]{revtex4}

\usepackage{amsmath} 
\usepackage{amssymb} 
\usepackage{epsfig}

\newcommand{\half}{\text{$\frac{1}{2}$}} 
\newcommand{\Tr}{\text{Tr}}
\newcommand{\p}{\partial} 
\newcommand{\bpsi}{\bar{\psi}} 
\newcommand{\bphi}{\bar{\phi}}
\newcommand{\dd}{\hbox{d}}

\begin{document}

\title{Quantitative Phase Diagrams of Branching and Annihilating Random Walks}

\author{L\'eonie Canet}
\affiliation{Laboratoire de Physique Th\'eorique et Hautes \'Energies, Universit\'{e}s Paris VI Pierre et Marie Curie, Paris VII Denis Diderot, 2 place Jussieu, 75251 Paris cedex 05, France}
\author{Hugues Chat\'e}
\affiliation{CEA-Service de Physique de l'\'Etat Condens\'e, Centre d'\'Etudes de Saclay, 91191 Gif-sur-Yvette, France}
\author{Bertrand Delamotte}
\affiliation{Laboratoire de Physique Th\'eorique et Hautes \'Energies, Universit\'{e}s Paris VI Pierre et Marie Curie, Paris VII Denis Diderot, 2 place Jussieu, 75251 Paris cedex 05, France}

\begin{abstract}
We demonstrate the full power of nonperturbative 
renormalisation group methods for nonequilibrium situations
by calculating the {\it quantitative} phase diagrams of simple 
branching and annihilating random walks and checking these results
against careful numerical simulations. Specifically, we show,
for the $2A\to\varnothing$, $A\to 2A$ case,
that an absorbing phase transition exists in dimensions $d=1$ to 6, and argue
that mean field theory is restored not in $d=3$, as suggested by previous 
analyses, but only in the limit $d \to \infty$.
\end{abstract}
\pacs{05.10.-a, 64.60.Ak, 64.60.Ht, 82.20.-w}

\date{\today}

\maketitle

The non-universal properties of continuous phase transitions, 
both at and out of thermal equilibrium, 
are much more difficult to determine than universal quantities
such as scaling exponents. 
The latter are generally accessible, even for strongly coupled systems, through perturbative calculations
near critical dimensions,
since renormalisation group (RG) transformations can then bring them 
in the vicinity of fixed points
 corresponding to weakly-coupled regimes.
On the other hand, non-universal properties, such as phase diagrams,
depend on the whole RG flow, which must be controlled to keep track 
of all the microscopic information. Such calculations are thus
genuinely non-perturbative.
This is usually tempered by the fact that the  mean-field (or one-loop)
approximation seems to capture semi-quantitatively the relevant 
non-universal physics of most equilibrium systems. 
Hence the common wisdom that trying to account
for fluctuations is needless since they are not expected to  
qualitatively alter phase diagrams.  
In this Letter, performing nonperturbative renormalization group (NPRG)
calculations checked against numerical simulations, we show that
this belief is drastically infirmed for a simple and widely studied 
nonequilibrium model, whose universal properties are 
otherwise well-understood.

We treat the case of branching and annihilating random walks 
(BARW) where particles of a single species $A$
diffuse at rate $D$ and undergo the reactions $A \xrightarrow{\sigma} 2A$, 
$2A\xrightarrow{\lambda}\varnothing$.
Whereas mean-field rate equations  predict that for any nonzero 
$\sigma$, the system always reaches an active steady state, 
early simulations \cite{tretyakov92,jensen93} in one and two space dimensions
have revealed the existence of a continuous transition
to the empty absorbing state, with scaling exponents characteristic of 
the directed percolation (DP) universality class \cite{hinrichsen00}.
Cardy and T\"auber \cite{cardy96} confirmed, 
through one-loop RG calculations, that fluctuations can 
indeed induce, for any nonzero annihilation rate $\lambda$, a 
dynamical DP transition in dimensions $d \le 2$.
Their analysis also suggested that mean field theory is recovered 
for $d > 2$, implying in particular that a transition can no longer 
occur in $d=3$, in agreement with numerical results by Takayasu
and Tretyakov \cite{tretyakov92}.

Below we determine the {\it quantitative} phase diagrams of 
BARW $A \xrightarrow{\sigma} 2A$, $2A\xrightarrow{\lambda}\varnothing$
in dimensions 1 to 6, from which it follows that 
the one-loop phase diagrams are qualitatively wrong above $d=2$, 
although perturbative RG does lead to the correct universal 
critical exponents.
Our approach uses and demonstrates the full 
power of the NPRG method, which allows to calculate
---in addition to the universal DP-class properties---
 non-universal quantities even beyond critical dimensions and weak coupling regimes.
All our results are confirmed, at the quantitative level, by
numerical simulations carefully tailored to approach, in a controlled
way, the continuous time limit where the analysis is performed.
Specifically, we show that DP-class transitions occur
at any nonzero $\sigma/D$ in all dimensions from 1 to 6,
but beyond a minimal threshold $(\lambda/D)_{\rm th}$ for $d \ge 3$.
The transition line drifts with increasing dimension,
and $(\lambda/D)_{\rm th}$ seems to grow linearly with $d$. This suggests
that the transition occurs for all $d$ and
that the mean field picture is realized only in infinite dimension.
Finally we show that the slope of the transition lines can 
be inferred from the master equation of a simple two-site model. 

We first briefly review the field theoretical formulation of BARW, 
and the nonequilibrium implementation of NPRG methods.
 The stochastic dynamics of BARW is described by a
master equation, which governs the time evolution of the 
configurational probability $P(\{n_i\},t)$. For instance the 
contribution of the birth and death processes at site $i$ reads:
\begin{eqnarray}
{\rm d} P(n_i,t)  = {\sigma} \: {\rm d}t 
\big[ (n_i \!-\! 1)P(n_i \!-\! 1,t) - n_i P(n_i,t)\big]+  \nonumber \\
 {\lambda} \: {\rm d}t \big[ (n_i \!+\! 2)(n_i \!+\! 1)P(n_i \!+\! 2,t) -  
n_i(n_i \!-\! 1)P(n_i,t)\big] 
\label{mastereq}
\end{eqnarray}
This master equation can be turned into a path integral representation 
in terms of two fields $\phi$ and  $\bphi$ \cite{doi76}.
Following \cite{cardy96}, 
the generating functional of the correlation functions reads
${\cal Z}[\phi,\bphi] = \int {\cal D}\phi \, {\cal D}\bphi \, 
\exp(-S[\phi,\bphi])$ with
\begin{eqnarray}
S[\phi,\bphi] & = &\int \dd^d r \, \dd t
\, \big[\bphi(\p_t - D \nabla^2)\phi  - \sigma\phi\bphi
\nonumber \\
& & + \sqrt{2\lambda\sigma}(\bphi\phi^2-\phi\bphi^2) + 
\lambda\phi^2\bphi^2\big] \;.
 \label{dp}
\end{eqnarray}
This field theory can be then investigated using the NPRG method.  
This formalism --- the effective average action
method \cite{tetradis94} --- is a continuous implementation, 
on successive momentum shells, of Wilson's block-spin concept.
It consists in building a sequence of running effective actions 
$\Gamma_k$ in which only fluctuations with momenta larger than the  
 running  momentum scale $k$ are averaged. 
At the scale $k=\Lambda$ ($\Lambda^{-1}$ corresponding to the 
underlying  lattice spacing), no  fluctuation is yet  taken into account
and $\Gamma_{k=\Lambda}$ coincides with the microscopic action $S$. 
At $k=0$, all  fluctuations are integrated out and $\Gamma_{k=0}$ 
encompasses the macroscopic properties of the system as it generates 
the one particle-irreducible correlation functions. $\Gamma_{k=0}$ is the analogue of the 
Gibbs free energy $\Gamma$ at thermal equilibrium. $\Gamma_k$ thus 
interpolates between the  microscopic action  and the effective action. 
The outcome of this procedure is twofold. First, it constitutes a 
RG method in the usual sense, in that the universal properties of 
critical phenomena can be derived from the behaviour of the flow 
in the vicinity of fixed points. On the other hand, since the NPRG 
flow can relate microscopic quantities to the large-scale behaviour 
of the system, this procedure embodies a calculation of the effective 
action associated with specific microscopic models. It thus enables to 
compute nonuniversal quantities, which depend on the  microscopic 
definition of the system. In this sense it radically differs from 
perturbative RG which loses memory of the microscopic details. 
Moreover, the flow equation of $\Gamma_k$ is nonperturbative, and as 
the approximations used do not rely on the smallness of a parameter, 
they are not confined to weak-coupling regimes or to the vicinity of 
critical dimensions.

\begin{figure}[ht]
\includegraphics[height=86mm,angle=-90]{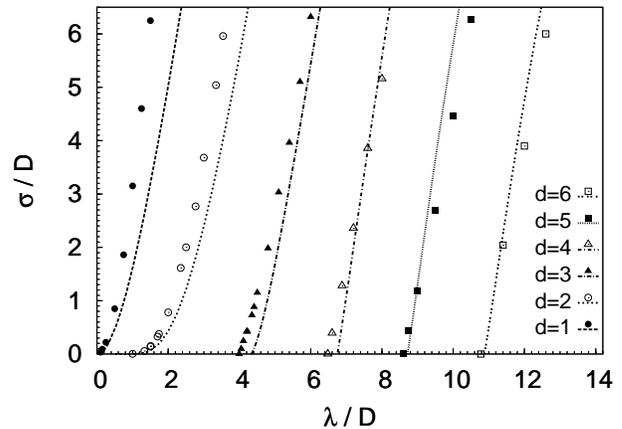}
\caption{Phase diagrams of BARW 
$A \xrightarrow{\sigma} 2A$, $2A\xrightarrow{\lambda}\varnothing$
in dimensions 1 to 6. Lines present NPRG results, rescaled as explained 
in the text. Symbols follow from numerical simulations. 
For each dimension, the active phase lies on the left of the transition 
line, the absorbing phase on the right.}
\label{diag}
\end{figure}

We now briefly sketch the construction of the effective action 
$\Gamma_k$ that only includes large momentum fluctuations, and  
give its nonequilibrium NPRG flow \cite{canet04} 
(see \cite{tetradis94,berges02} for a detailed derivation at 
thermal equilibrium). 
The low and high momentum fluctuation modes are separated by a 
scale dependent mass-like term  
$\Delta S_k(\phi,\bphi) = \int_q \Phi(-q)\hat{R}_k(q) \Phi^T(q)$ 
where $\Phi =(\bphi,\phi)$, and $\hat{R}_k(q)$ is a symmetric 
matrix with zeros on its diagonal, and a cutoff function  $R_k(q)$ 
off its diagonal.
This cutoff acts as an effective mass  $R_k(q)\sim k^2$  for $q\ll k$  
that suppresses the propagation of the low momentum modes,
while  not altering the high momentum ones,  $R_k(q)\to 0$ for $q\gg k$. 
In this work, we use $R_k(q)= (k^2-q^2)\theta(k^2-q^2)$  \cite{litim01}, 
which leads to simple analytical expressions ($\theta(x)$ denoting the step function).
 $\Gamma_k$ is obtained by a Legendre transform of log${\cal Z}_k$, 
modified by an additional contribution $\Delta S_k(\psi,\bpsi)$ [$\bpsi$ 
and $\psi$ being the functional derivatives of  log${\cal Z}_k$ w.r.t. 
sources], in order to ensure that the proper limits are recovered at the 
scales $k=0$ and $k=\Lambda$.
An exact RG equation for the flow of $\Gamma_k$ is easily derived and 
underlies the nonequilibrium NPRG formalism \cite{canet04}:
\begin{equation}
\p_k \Gamma_k (\psi,\bpsi)=\half \Tr \left\{ [\hat{\Gamma}_k^{(2)}+\hat{R}_k]^{-1}\p_k \hat{R}_k\right\},
\label{flow}
\end{equation}
where $\hat{\Gamma}_k^{(2)}$ is the field-dependent matrix of the second 
functional derivatives of $\Gamma_k$, and Tr stands for matrix trace over 
internal indices and integration over the internal momentum and frequency.
Eq. (\ref{flow}) is a functional equation  which obviously cannot be 
solved exactly. To handle it, one has to {\it truncate} $\Gamma_k$. 
We here implement a standard truncation consisting in  the leading order 
of a time and space derivative expansion of  $\Gamma_k$,  which  reads:
\begin{equation}
\Gamma_k(\psi,\bpsi)=\int\dd^d r \, \dd t\, [Z_k \bpsi (\p_t - D_k \nabla^2)\psi\\
 + V_k(\psi,\bpsi)].
\label{GammaDP}
\end{equation}
We emphasize that the full functional dependence of  $V_k(\psi,\bpsi)$ 
is dealt with and this truncation hence embodies an accurate description 
of the steady and uniform configurations of the system,  as supported by 
all previous studies at thermal equilibrium.  
We refer to \cite{berges02,morris98,canet03a,canet04} for detailed 
discussions of the approximation schemes. 
The flow equations for $Z_k$, $D_k$ and $V_k(\psi,\bpsi)$ are given in 
\cite{canet04}. They allow us to determine the phase diagrams of BARW 
$A \xrightarrow{\sigma} 2A$, $2A\xrightarrow{\lambda}\varnothing$
in any dimension.
For this, we numerically integrate the flow equations from an arbitrary 
initial ultra-violet scale $k =\Lambda$ where $\Gamma_{k=\Lambda}$ 
identifies with the microscopic action~(\ref{dp}) for given $\lambda$, 
$\sigma$ and $D$, and we determine the phase of the system 
(at the final scale $k=0$) ensuing from this initial condition.
Figure~\ref{diag} (lines) shows the resulting phase boundaries for
$d=1$ to 6.
 
Before commenting on the phase diagrams, we present numerical results
which fully confirm them. No existing simulations are available 
for a quantitative comparison. Worse, in $d=3$, no transition 
was found \cite{tretyakov92}. 
In these simulations the rates $\sigma$, $\lambda$, and $D$ were 
parameterized by a single free variable, a drastic
constraint likely to prevent them from reaching the absorbing state.  
Moreover, in \cite{tretyakov92}, a strict occupation restriction was
enforced (``fermionic'' model), at odds with the analytical context.
Here, we adapt the efficient models introduced recently in \cite{chate02}
to simulate arbitrary BARW processes without strict occupation restriction.
We first notice that modifying the models of \cite{chate02} to study
$A \xrightarrow{\sigma} 2A$, $2A\xrightarrow{\lambda}\varnothing$
for $d\ge 3$ readily provides evidence of the existence of DP-class 
transitions (not shown).
This does not allow, however, a quantitative comparison with our analytical
results. Reaching this aim requires to resort to a time 
discretisation $\Delta t$ of the master equation which reproduces 
as accurately as possible the continuous time evolution. 
A proper discretisation is achieved in the limit of $\Delta t$ small enough
to ensure that $\Delta P$ also remains infinitesimal so that the 
 obtained critical rates remain invariant under rescaling $\Delta t$. 
We proceed as follows. At each time step, all sites $i$ undergo 
a parallel update. The on-site reactions are ruled by the three 
independent rates $\sigma$, $D$ and $\lambda$: each of the $n_i$ particles 
is tested for branching and for diffusion to a nearest-neighbor 
with respective probabilities $\sigma \Delta t$ and $D \Delta t$ 
for each neighbor. Simultaneously, each of the $n_i(n_i-1)/2$ possible 
pairs are candidates for annihilation with probability $2 \lambda \Delta t$ 
following Eq.~(\ref{mastereq}). The occupation number $n_i$ is then
updated according to the successful trials and the diffusion moves
are implemented. 
We observe that the alteration of critical rates induced  by changing  
$\Delta t$ to $\Delta t/10$ generally does not exceed a few percent 
as long as the probabilities $\sigma \Delta t$, $2 \lambda \Delta t$ and 
$D \Delta t$ remain smaller than typically $10^{-3}$. In this limit,
draws such that the number of annihilated particles would be larger
than $n_i$ are exceedingly rare (and rejected anyway).
It is only in these rather inefficient-looking regimes that we
can ensure to approach the continuous time limit. 
The code remains however powerful because 
only (the typically few remaining) non-empty sites are visited.
The critical rates are determined by tracking the decay of the
total population starting from fully-occupied initial conditions.
The critical point is characterized by an algebraic decay with
the DP exponent $\delta=\beta z/\nu$, separating saturation (supercritical
regime) from (quasi-) exponential decay. To obtain the typical accuracy
of $2\%$ of the results presented in Fig.~\ref{diag} (symbols), 
system sizes up to $2^{24}$ sites,
and simulation times of up to $10^8$ steps were necessary.

\begin{figure}[ht]
\includegraphics[height=86mm,angle=-90]{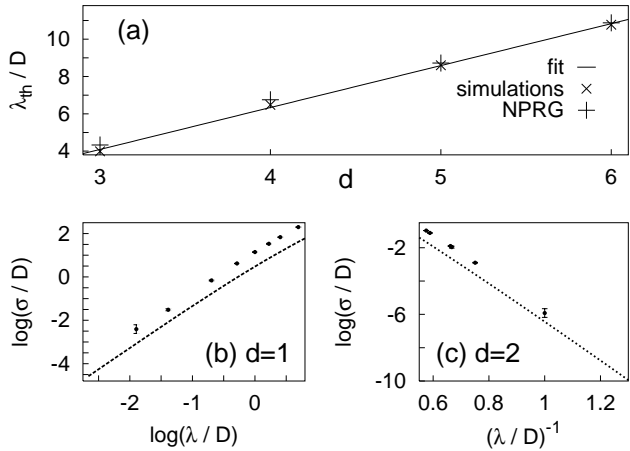}
\caption{(a): Evolution of the thresholds $(\lambda/D)_{\hbox{\footnotesize th}}$ with the dimension; (b) and (c): log plots of the transition line in the vicinity of  the origin in $d=1$ and $d=2$ with error bars.}
\label{thres}
\end{figure}

Still, the comparison between the analytical and numerical phase diagrams
requires the prior calibration of the axes $\lambda/D$ and $\sigma/D$.
Indeed, as simulations implement a discrete lattice version of the master 
equation, the ``numerical'' rates differ from the analytical 
ones by a spatial continuum limit. We therefore introduce an overall 
offset parameter $C$ which can be interpreted as the ratio 
$\Lambda_0/\Lambda$ of the corresponding underlying microscopic scales 
---respectively lattice $\Lambda_0$ and ultra-violet $\Lambda$. 
The ratios $\lambda/D$ and $\sigma/D$ are therefore corrected 
with regards to their scaling dimensions by factors $C^{2-d}$ and $C^2$ 
(see Eq.~(\ref{dp})). The unique offset parameter $C$ is fixed by fitting
 the NPRG thresholds $C^{2-d} (\lambda/D)_{\rm th}(d)$ to the numerical
 ones. This produces a very accurate match (Fig.~\ref{thres}(a)). 
The resulting rescaled analytical full transition lines 
then also closely match their numerical counterparts on the range of rates considered in all dimensions 
from 1 to 6, as displayed in Fig~\ref{diag}. 
[We have checked that fitting other quantities to fix $C$ only mildly changes its value, which
does not spoiled the agreement between the numerical and analytical diagrams.]

%
%

%
%

We now comment on the phase diagrams. Their main feature
is that the phase transition exists in all probed dimensions, thus qualitatively invalidating
 the one-loop picture. 
For $d \geq 3$, the transition curves are almost parallel straight lines, 
crossing the $\lambda/D$ axis at a nonzero threshold value 
$(\lambda/D)_{\rm th}$.
Up to $d=6$, this threshold grows linearly with $d$ at a rate 
$2.248\pm0.058$, as  extracted from a linear fit (see Fig~\ref{thres}(a)). 
Further theoretical calculations confirm that 
$(\lambda/D)_{\rm th}(d)$  is nearly linear at least up to $d=10$. 
This suggests that  $(\lambda/D)_{\rm th}$  becomes infinite  in the limit 
$d \to \infty$, so that only the active phase remains in this limit. 
In other words, the mean field phase diagram seems to be recovered
 only at infinite dimension, that is neither in $d=3$ (one-loop) nor in $d=4$ (upper critical dimension).
Below $d=3$, the threshold vanishes. The approach of the 
transition curve to the origin is quadratic in $d=1$, and exponential 
in $d=2$ with a coefficient $11.86\pm 0.02$ analytically, and $11.67\pm 0.15$ 
numerically (Fig.~\ref{thres}(b) and (c)). 
This is in close agreement with the coefficient $4\pi$ 
predicted by perturbative RG (see \cite{cardy96} and \cite{canet04}). 

%
\begin{figure}[ht]
\includegraphics[height=86mm,angle=-90]{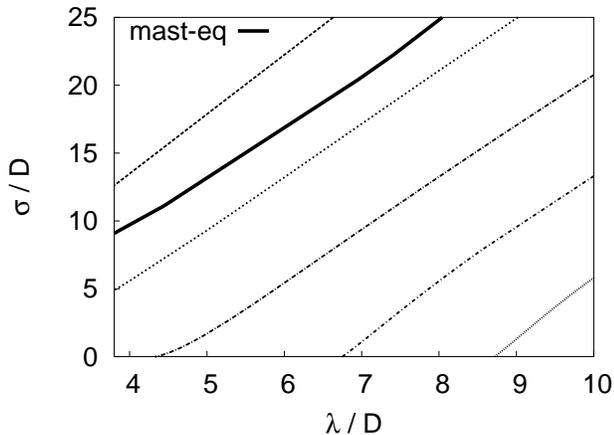}
\caption{Dashed lines: NPRG phase diagrams in $d=1$ to $d=5$ 
(from left to right); thick solid line: result from the two-site model.}
\label{master}
\end{figure}
%

We finally give a simple argument supporting our finding that in the weak
diffusion regime the phase boundaries are almost straight lines.
 Indeed, an effective transition line 
can be estimated from the master equation itself in the limit of
large $\sigma/D$ and $\lambda/D$.
If $D=0$ exactly, sites are 
decoupled and the time evolution of each site can be considered 
independently. The evolution of the probability $P(n,t)$ at a given site 
is governed in this limit by the infinite set of coupled differential 
equations (\ref{mastereq}). One can readily prove from them that the unique 
stationary solution is $\{P(n=0)=1$, $P(n \ge 1)=0\}$, which corresponds 
to the absorbing state. For $\sigma/\lambda \lesssim 10$, we have checked, 
by numerically integrating the master equations truncated to $n = 100$, 
that this state is the unique attractor reached at large times. 
(Including  higher occupation number equations does not change this result.)
Thus, for $D=0$, the system always ends up in the 
absorbing phase at least up to $\sigma/\lambda \lesssim 10$ and probably 
for any finite $\sigma/\lambda$, with a relaxation time $\tau$ growing 
with $\sigma/\lambda$. We now argue that the absorbing 
state remains stable when a small diffusion is allowed. Qualitatively, 
we expect this  to hold true as long as the diffusion time $(2 d D)^{-1}$
($2d$ standing for the number of neighbors), 
remains much larger than $\tau$ since then
the particles on one site die before diffusing. 
To make this statement more quantitative, we consider the following model. 

Near criticality, the system is very diluted, so we focus on an 
isolated single particle at site $I$, and 
model the rest of the lattice by a single, initially empty, neighboring 
site $J$. We numerically integrate a ($30 \times 30$) truncation of the 
master equation for $P(\{n_I,n_J\},t)$. Again, the system always reaches 
the absorbing phase, but we can assume that an active state can be sustained
if the particle at site $I$ manages to multiply and spread out to 
the neighbouring site before it dies out. We hence study the average 
occupation numbers  $\bar{n}_I(t)$ and  $\bar{n}_J(t)$. At given $\lambda$ 
and $\sigma$, as long as $D=0$,  $\bar{n}_J(t)=0$. When $D$ is increased,  
$\bar{n}_J(t)$ starts to grow and reaches a maximum at $t=t_m$ before 
eventually vanishing. This leads us to set up the following criterion: 
the absorbing state is supposed to be destabilized if $\bar{n}_J(t_m)$  
reaches one (while $\bar{n}_I(t_m) \ge 1$),  which defines a critical 
diffusion rate $D_c$. Probing various $\sigma/\lambda$ yields a 
nearly-straight transition line with approximately the observed
constant slope (Fig.~\ref{master}). 
This argument, as simple as it is, is by no means a rigorous proof,
but provides further support to the existence of an absorbing state 
in the weak diffusion regime, in all finite dimensions.

In summary, we have provided a clear and coherent picture of the 
quantitative phase diagrams of BARW
$A \xrightarrow{\sigma} 2A$, $2A\xrightarrow{\lambda}\varnothing$
for $d=1$ to 6.
We conclude that DP-class transitions
probably exist in all space dimensions, and  only occur
beyond a minimum decay rate $(\lambda/D)_{\rm th}$ for $d>2$. Note 
that it is the very existence of this threshold which
dooms any perturbative approach.
The close agreement between our simulations and 
our analytical results firmly roots the ability of the 
NPRG method to probe with great accuracy the nonuniversal 
behaviour of a specific model. 
This offers a promising means of investigation of other reaction-diffusion 
processes. Different types of BARW are obvious candidates \cite{cardy96},
and future work will be devoted to explore the so-called PCPD
(``pair-contact process with diffusion'')
$2A \to 3A$, $2A\to\varnothing$ \cite{henkel04}.

\begin{acknowledgments}
B.D. wishes to thank G. Oshanin for fruitful discussions. The LPTHE is Unit\'e Mixte du CNRS, UMR 7589.
\end{acknowledgments}

\end{document}